\documentclass{aa}  
\usepackage{graphicx}
\usepackage{txfonts}
\usepackage{soul}
\usepackage[colorlinks=true, citecolor=blue]{hyperref}
\usepackage{natbib,twoopt}
\bibpunct{(}{)}{;}{a}{}{,}             
\usepackage{hyperref}
\hypersetup{
    colorlinks=true,
    linkcolor=blue,
    citecolor = blue,
    filecolor=magenta,
    urlcolor=blue}
\usepackage{booktabs}
\usepackage{lscape}
\usepackage{float}
\usepackage{ulem}
 \usepackage[utf8]{inputenc} 
 
\begin{document}

\authorrunning{Origlia et al.}
\titlerunning{X-shooter spectroscopy of giant stars in the nuclear star cluster}

\title{X-shooter spectroscopy of giant stars in the nuclear star cluster
\thanks{Based on  observations at the Very Large Telescope of the European Southern Observatory at Cerro Paranal (Chile), under program 099.D-0258 (PI L. Origlia).}}

\author{L. Origlia\inst{1},  
    C. Fanelli\inst{1}, 
    A. Mucciarelli\inst{2}\fnmsep\inst{1},
    R. M. Rich\inst{3},
    M. Schultheis\inst{4},
    N. Ryde\inst{5},
    B. Thorsbro\inst{4},
    T. Fritz\inst{6}
}

\institute{INAF, Osservatorio di Astrofisica e Scienza dello Spazio di Bologna, Via Gobetti 93/3, I-40129 Bologna, Italy\\
\email{livia.origlia@inaf.it}
         \and
         Dipartimento di Fisica e Astronomia “Augusto Righi”, Alma Mater Studiorum, Universitá di Bologna, Via Gobetti 93/2, 40129 Bologna, Italy
         \and
         Department of Physics and Astronomy, UCLA, 430 Portola Plaza, Box 951547, Los Angeles, CA 90095-1547, USA
         \and
         Université Côte d’Azur, Observatoire de la Côte d’Azur, CNRS, Laboratoire Lagrange, 06000 Nice, France
         \and
         Division of Astrophysics, Department of Physics, Lund University, Box 118, 221 00 Lund, Sweden
         \and
         Fraunhofer IOSB (Institute of Optronics, System Technologies and Image Exploitation), Department Object Recognition, Gutleuthausstr. 1, 76275 Ettlingen, Germany
         }

\abstract{
Near-IR spectra of a homogeneous sample of 15 red giants in the Galactic nuclear star cluster, within 1.5 pc of SgrA*, have been acquired with X-shooter at the Very Large Telescope. 
From these spectra  line-of-sight radial velocities and chemical abundances of iron, carbon, oxygen and other alpha and light elements have been derived.
By combining radial velocities from this study and proper motions from the literature we computed the orbits of these stars, finding that they have been confined within 12 pc from SgrA* for their entire lifetime, thus strongly suggesting an in-situ formation and evolution.
Iron abundances between half and twice solar, about solar-scaled values within $\pm$0.1 dex for $\alpha$-elements, Ti and V, some enhancement of [Na/Fe], [K/Fe] and [Al/Fe] and some depletion of [C/Fe] with respect to the solar ratios have been measured. The inferred chemical abundance distributions are consistent with a formation from a gas enriched by both type II and type I SNe over a prolonged timescale, closely matching those of the metal-rich populations of the inner bulge field and other complex stellar systems like Terzan~5 and Liller~1.
}
\maketitle

\section{Introduction} 

\begin{table*}
\caption{Coordinates, photometry, proper motions, heliocentric radial velocities (RVs), effective temperatures and orbital parameters for the observed NSC stars.}
\label{tab1}
\centering
\small
\setlength{\tabcolsep}{5.5pt}
\begin{tabular}{|ccccccccc|cccc|}
\hline\hline 
ID & RA & Dec & H & K & PM$_{\rm RA}$ & ePM$_{\rm RA}$ & PM$_{\rm Dec}$ & ePM$_{\rm Dec}$ & RV(hel) & T$_{\rm eff}$ & r$_{\rm apo}$ & z$_{\rm max}$ \\
 & [deg] & [deg] & [mag] & [mag] & [mas yr$^{-1}$] & [mas yr$^{-1}$] & [mas yr$^{-1}$] & [mas yr$^{-1}$] & [km s$^{-1}$] & [K] & [pc] & [pc] \\
\hline
\multicolumn{9}{|c|}{\citealt{fritz16}} & \multicolumn{4}{|c|}{This study}  \\
\hline
5657 & 266.422370 & -29.003613 & 13.04 & 10.85 & -3.705 & 0.149 & -0.263 & 0.140 & -68.6 & 3500 &  10.0 & 6.9 \\
5726 & 266.424149 & -29.006863 & 13.45 & 10.73 &  2.412 & 0.133 &  2.754 & 0.133 &  81.3 & 3550 &  10.5 & 7.3 \\
6157 & 266.412429 & -29.002155 & 13.63 & 10.61 &  0.131 & 0.237 & -2.948 & 0.373 & -16.6 & 3650 &   9.9 & 5.7 \\
6508 & 266.422112 & -29.002327 & 12.48 & 10.60 &  0.237 & 0.247 &  2.274 & 0.321 & -62.2 & 3500 &   9.8 & 6.8 \\
6621 & 266.412117 & -29.013822 & 13.09 & 10.53 & -0.310 & 0.227 &  1.537 & 0.227 &  -8.2 & 3500 &  11.2 & 6.6 \\
6993 & 266.425308 & -29.009852 & 13.33 & 11.27 &  1.652 & 0.218 &  1.108 & 0.218 & -28.3 & 3850 &  11.1 & 7.6 \\
7363 & 266.422722 & -29.001611 & 12.85 & 10.92 &  2.972 & 0.128 &  1.359 & 0.128 &  -1.5 & 3500 &   9.8 & 6.8 \\
7537 & 266.424362 & -29.012805 & 13.01 & 10.51 &  3.022 & 0.227 &  3.009 & 0.227 & -21.9 & 3500 &  11.5 & 7.8 \\
7651 & 266.414566 & -28.999649 & 13.13 & 10.90 & -3.291 & 1.953 & -1.328 & 1.953 & 126.7 & 3500 &   9.3 & 5.8 \\
7680 & 266.425096 & -29.012182 & 13.18 & 10.68 & -0.655 & 0.227 &  2.618 & 0.227 &  14.8 & 3400 &  11.3 & 7.9 \\
7699 & 266.409342 & -29.002460 & 13.02 & 10.79 &  1.125 & 0.227 &  2.279 & 0.227 &  38.3 & 3800 &   9.8 & 5.5 \\
8116 & 266.414979 & -28.998944 & 13.16 & 10.65 &  3.683 & 1.870 &  2.867 & 1.870 &-123.9 & 3400 &   9.2 & 5.8 \\
8323 & 266.412269 & -29.016230 & 12.57 & 10.76 &  0.755 & 0.227 &  1.600 & 0.227 & -55.4 & 3750 &  11.6 & 7.0 \\
8684 & 266.407325 & -29.002588 & 13.26 & 10.73 & -4.096 & 0.236 & -0.963 & 0.236 &  44.1 & 3300 &   9.6 & 5.2 \\
9349 & 266.408588 & -29.016127 & 12.93 & 10.53 & -1.366 & 0.227 &  0.040 & 0.227 &  13.1 & 3550 &  11.6 & 7.2 \\
\hline\hline
\end{tabular}
\end{table*}

In the era of Gaia \citep{gaia_mission} and ground-based massive spectroscopic surveys of the old stellar populations in the Milky Way halo and bulge/disk at 4-8m class telescopes, 
radial velocities and chemical abundances for large samples of giant stars in the Galaxy field and star clusters are being obtained, providing fundamental information on the formation and chemical enrichment history of the various Galactic components.
A few other massive surveys are planned in the near-future by using the next generation of multi-object spectrographs with unprecedented large multiplexing \citep{cirasuolo16,4MOST19,WEAVE24}, to provide a comprehensive kinematic and chemical mapping of the stellar populations of the Milky Way.
Within the Galaxy however, the innermost nuclear region towards the Galactic Center (GC) remains poorly explored, despite its uniqueness and importance to understand e.g. co-evolution and feedbacks of the central black-hole and the surrounding host, mostly because of the extinction, that it is so severe that only observations at IR and radio wavelengths are possible.
APOGEE \citep{majewski17} and a few other small surveys at Keck, VLT and Gemini are making use of medium-high resolution red and near-IR spectroscopy to observe giant stars in the central hundreds pc, in order to get  chemical abundances of iron, iron-peak and  neutron-capture elements, CNO and some other alpha and light elements.
In particular, measurements of cool giants in the central 500 pc from the GC \citep[see, e.g.,][and references therein]{rich07,rich12,ryde16a,nandakumar18,schultheis19,schultheis20,feld20,fritz21,Nieu23} indicate iron abundances from 1/3 to twice solar, with a tail toward [Fe/H]$<$-0.5 dex and some alpha enhancement.
Measurements of red supergiants and some older giants close to the GC 
\citep[see, e.g.,][and references therein]{ramirez00,cunha07,davies09,ryde15,nandakumar18}
are consistent with peak metallicity around solar and low (if any) level of alpha-element enhancement. 

The Galactic nucleus discovered by \citet{BN68} has since been described with greater precision by \citet{morris96,launhardt02}. 
The early star formation and chemical enrichment history of the nucleus is expected to be somewhat unique. 
Indeed, the presence of the supermassive black hole (SMBH) should significantly affect star formation via tidal forces \citep[see, e.g.,][]{ghez07,genzel10} and stellar orbits within its sphere of influence, while outside of it the stellar populations should be more similar to those of the inner bulge/disk.
Star counts and kinematics \citep[][and references therein]{schodel14,feldmeier14,fritz16} reveal a central, compact structure, the so-called nuclear star cluster (NSC), with a half-light radius of 7 pc and a mass of about 2.5$\times 10^7$ M$_{\odot}$, that also hosts the supermassive black hole, SgrA$^\ast$. Such a NSC contains stars covering a wide range of ages and metallicities, suggesting a quasi-continuous star formation history.
It might be the surviving core of multiple globular clusters or other stellar systems that have migrated 
towards the center from outside the nuclear region \citep[see, e.g.,][]{tremaine75,capuzzo93,antonini13,gnedin14}, or  it could have formed in situ \citep[see, e.g.,][]{milosavljevic04,seth08,pflamm09}, with also the possible contribution of stars from the inner bulge and/or disk. 
Some metallicity gradient between the NSC and the nuclear stellar disc (NSD) \citep[see e.g.][]{feld22} has been also claimed.

In order to solve the puzzle of the NSC formation and evolution, it is of fundamental importance to determine individual abundances of iron, 
CNO, other alpha and light and heavy elements of old giants, likely members of the NSC, and compute suitable 
abundance ratios that are crucial to constrain star formation and chemical enrichment timescales.
K-band spectroscopy of giant stars in the central parsec of the NSC
\citep{do15,feldmeier17} provided overall metallicities spanning a wide range from about 1/10 to supersolar values. However, the limited spectral resolution (R$\sim$4-5000) of these observations prevented the measurement of individual chemical abundances and a proper chemical tagging of the old stellar population(s) in this environment.
\citet{ryde16b} provide some detailed chemical abundances of iron and a few alpha elements of a metal poor, 
alpha enhanced bright giant at $<$30pc from SgrA*.
A more comprehensive sample of giants stars in the NSC analyzed by \citet{rich17,tho20} indicate iron abundances around solar with a spread of $\sim$0.3 dex and some level of [Si/Fe] enhancement.
More recently, high resolution H and K band spectroscopy of a few, metal-rich cool giants in the NSC \citep{ryde25,nandakumar25} provides chemical abundances of iron-peak and alpha-elements that follow the trends of the metal-rich population in the inner bulge. 

Taking advantage of the X-shooter \citep{vernet11} spectrograph at the VLT with its medium resolution, simultaneous wide spectral coverage 
and high throughput in the near IR, we efficiently observed a sample of 15 stars likely members of the NSC and located within about 1.6 pc from the GC.
Observations and data reduction are presented in Sect.~\ref{obs}, spectral analysis is discussed in Sect.~\ref{sp}. Sect.~\ref{kin} and \ref{chem} present the results of our kinematic and chemical analyses, respectively, while Sect.~\ref{conc} reports some final considerations and conclusions.

\section{Observations and data reduction}
\label{obs}
We observed 15 giant stars, candidate members of the NSC, with X-shooter at VLT  under program 099.D-0258 (PI L. Origlia).
Because of the huge reddening towards the NSC, only the spectra acquired with the  NIR arm and the 0.6 arcsec slit (providing a resolution of R$\approx$8,000) in the 1.15-2.37 $\mu$m range, have been used for spectral analysis.

\begin{figure*}
    \centering
    \includegraphics[width=0.8\textwidth]{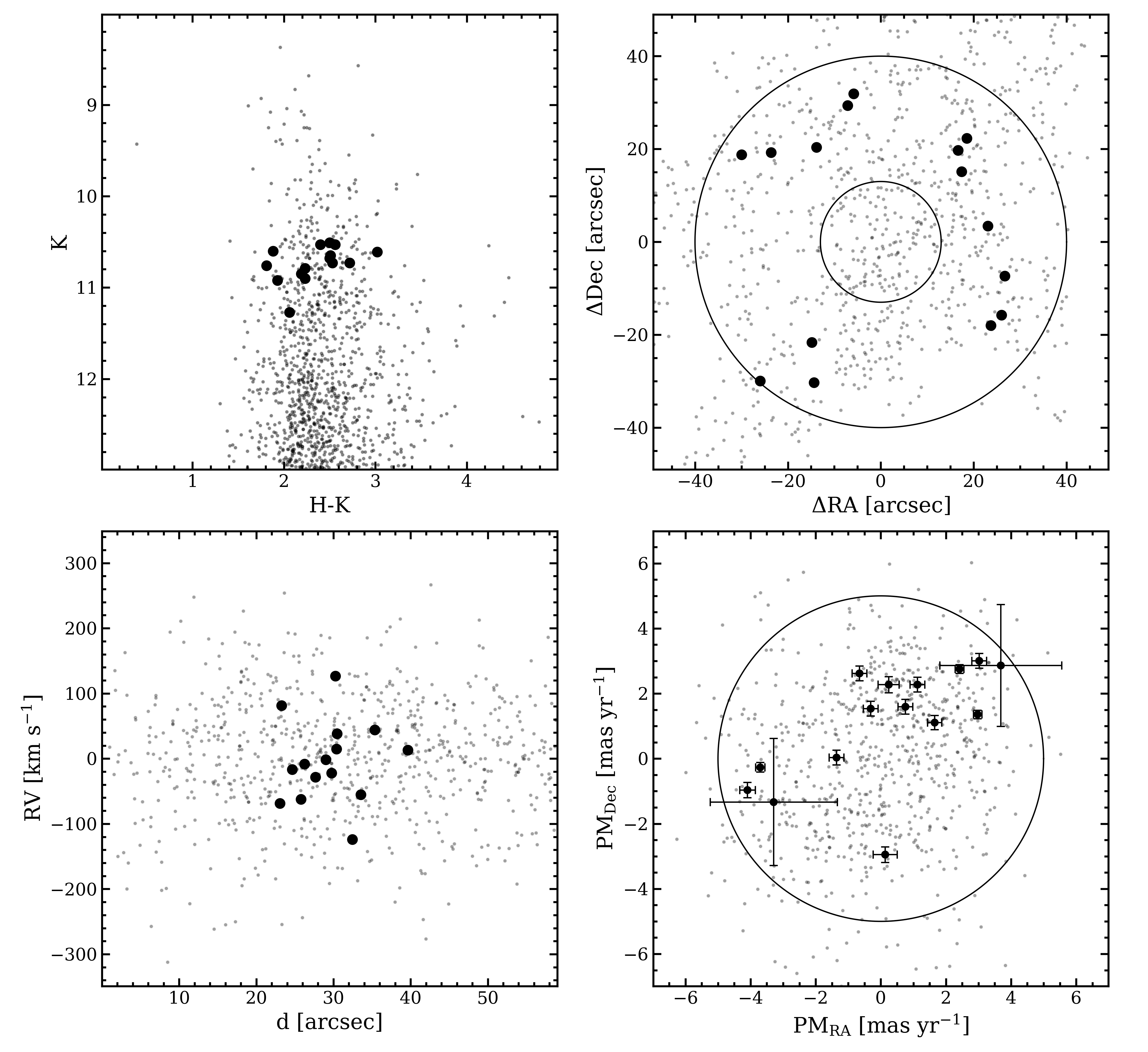}
    \caption{K-H,K color-magnitude diagram (top-left), RA-Dec map (top-right), heliocentric RVs as a function of the projected distance from SgrA$^*$ (bottom-left) and proper motions (bottom-right) for the stars toward the NSC (gray dots) measured by \citet{fritz16}.  Targets observed with X-shooter are indicated as black big dots. In the top-right panel the two big circles delimit the regions at projected distances on sky of 13 and 40 arcsec, corresponding to about 0.5 and 1.6 pc, respectively, from SgrA$^*$. In the bottom-right panel, the circle has a radius of 5 mas yr$^{-1}$.}
    \label{targets}
\end{figure*}
The targets have been extracted from the catalog of \citet{fritz16} by using the quoted H,K band photometry, RA and Dec coordinates and proper motions (see Table~\ref{tab1}) and applying the following selection criteria.\\
{\it i})~By using the observed K,H-K color-magnitude diagram (see Fig.~\ref{targets}, left panel), 
in order to maximize the probability of selecting candidate NSC member old giants below the Red Giant Branch (RGB) Tip on one hand, 
and to observe stars bright enough to get X-shooter spectra with sufficient signal-to-noise ratio in a reasonable amount of integration time on the other hand, we selected stars with K in the 10-12 magnitude range and very red H-K in the 1.8-3.2 color range to account for the huge absolute and differential reddening in that direction.\\
{\it ii)}~We selected stars located in an annular region between about 0.5 and 1.6 pc assuming a distance of 8.2 kpc ({\it i.e.} between 13 and 40 arcsec projected on sky) from the GC (see Fig.~\ref{targets}, middle panel), in order to sample 
the SMBH sphere of influence but avoiding the innermost crowded region at r$<$0.5~pc, where seeing limited observations are problematic.\\
{\it iii})~We selected stars with proper motions of $|\mu _{RA}|$ and $|\mu_{Dec}|$ $\leq$ 5 mas/yr (see Fig.~\ref{targets}, right panel), thus consistent with being NSC members.

Since the selected target stars  suffer an average visual extinction of about 30 mag, thus being too faint to be detected with the X-shooter acquisition camera that works in the R band, and hence to be directly centered in the slit, blind offset from a brighter, nearby reference star has been used.
Acquisition of X-shooter spectra has been then performed by nodding on slit with a typical throw of a few arcsec, for a proper background and detector subtraction. An O-star spectrum observed during the same night has been used to check and eventually remove telluric features. Typical total on-source exposure times ranged from 30 to 60 min, depending on the target brightness.\\

\begin{figure*}
    \centering
    \includegraphics[width=\textwidth]{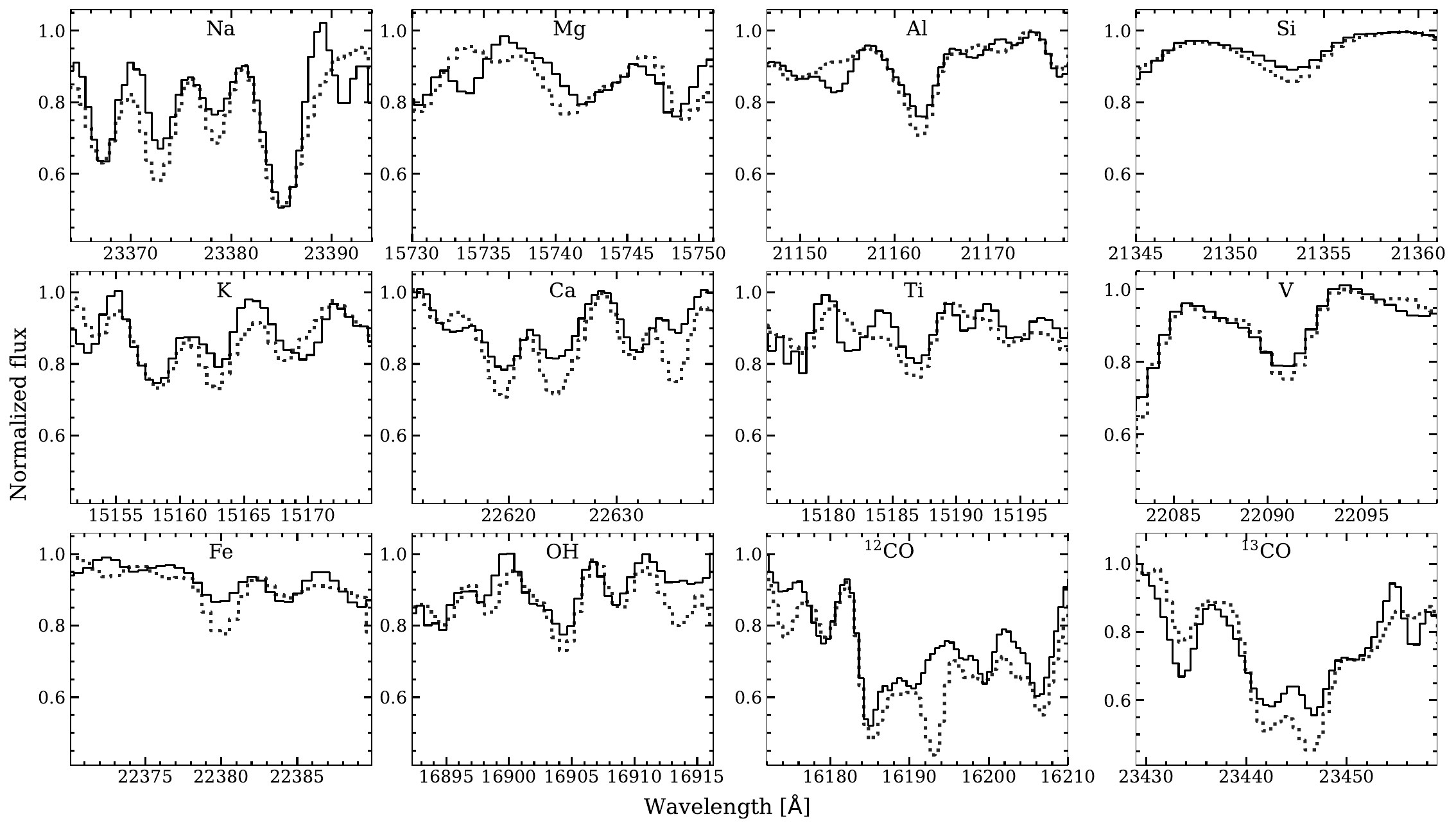}
    \caption{Portions of normalized, rest-frame X-shooter NIR spectra around some atomic and molecular lines of interest for stars 5726 (solid) and 7363 (dotted) with similar stellar parameters and different metallicities.}
    \label{spectra}
\end{figure*}

The reduction of the X-shooter NIR spectra has been performed by using the ESO X-shooter pipeline to obtain background subtracted, flatfield corrected, rectified and wavelength calibrated 2D spectra. 
The 1D spectrum extraction was performed manually in order to visually inspect and optimize the location and extension on the detector of the target-sky pair of spectra corresponding to the A and B positions along the slit. 
Due to the severe extinction towards the NSC, only the portion of the X-shooter spectra covering the H and K bands have enough signal to be effectively used for the subsequent science analysis. The signal to noise ratio of the final, extracted spectra at $\lambda >$1.5 $\mu$m is always $>$30. 
Fig.~\ref{spectra} shows an example of observed spectra around some atomic and molecular lines of interest for two stars with similar stellar parameters and different metallicities.

\section{Spectral analysis}
\label{sp}

The R=8,000 spectral resolution and the simultaneous wide NIR spectral coverage of the calibrated X-shooter spectra, coupled with spectral synthesis and cross-correlation techniques,  enabled to estimate  stellar parameters, radial velocities and chemical abundances of individual elements for the 15 observed stars.

Absolute and differential reddening toward the NSC is so severe and affected by non-negligible uncertainties to prevent any reliable photometric estimate of the stellar temperature. Hence, as already done in previous studies of the NSC \citep[see e.g.][]{rich17,feldmeier17, ryde25, nandakumar25} stellar temperatures have been derived from the (2-0) and (3-1) $^{12}$CO bandheads in the K-band and the empirical relations of \citet{schultheis16}, while gravity log~g=0.5 and  microturbulence of 2 km~s$^{-1}$, typical of luminous bulge and GC giant stars with similar low temperatures \citep[see e.g.][]{rich12,rich17}, were assumed for all the observed stars. 
The derived temperatures in the 3200-3800 K range have been also cross-checked against OH lines, in order to obtain a simultaneous best-fit of all the molecular CO and OH transitions.

Uncertainties in the derived stellar parameters are mostly systematic and for the observed cool giants near the RGB Tip we estimated values of $\pm$100 K in temperatures, $\pm$0.3 dex in log(g) and $\pm$0.2 km s$^{-1}$ in microturbulence. The overall impact of these uncertainties in the derived [X/H] abundances and [X/Fe] abundance ratios is always within 0.15 dex, normally within 0.1 dex, and  they have mostly the effect of rigidly shifting the chemical distributions. However, since these shifts turn out to be relatively small (if any), they do not significantly affect the overall appearance of the inferred distributions.

Heliocentric radial velocities (RVs) have been obtained via cross-correlation of the observed spectra with suitable templates whose stellar parameters closely match those of the target stars in the 2.29-2.37 $\mu$m spectral window in the K-band, which includes several CO lines and bandheads.
The RV uncertainty was estimated using the full width at half maximum of the cross-correlation function combined with the signal-to-noise ratio of the observed spectra, resulting in an error of $\approx$2 km~s$^{-1}$ \citep{tonry79}.
The derived heliocentric RVs were also cross-checked and validated with the atomic and molecular lines in the H-band.

\begin{table*}
\caption{Chemical abundances$^*$ and corresponding errors$^{**}$ for the observed NSC stars.}
\label{tab2}
\small
\setlength{\tabcolsep}{4pt}
\begin{tabular}{|c|c|c|c|c|c|c|c|c|c|c|c|c|}
\hline\hline
ID & [Fe/H] & [C/H] & [O/H] & [Na/H] & [Mg/H] & [Al/H] & [Si/H] & [K/H] & [Ca/H] & [Ti/H] & [V/H] & $^{12}$C/$^{13}$C\\
&    [dex]    & [dex]    & [dex]    & [dex]    & [dex]    & [dex]    &  [dex]    &  [dex]    &  [dex]    & [dex]  &  [dex] &  \\
\hline
6508 & +0.33 & +0.02 & +0.22 & +0.76 & +0.29 & +0.30 & +0.31 & +0.22 & +0.27 & +0.36 & +0.25 &  7 \\
     & $\pm$0.05 (7) & $\pm$0.04 (4) & $\pm$0.08 (4) & $\pm$0.02 (2) & $\pm$0.04 (2) & $\pm$0.10 (1) & $\pm$0.06 (3) & $\pm$0.04 (2) & $\pm$0.08 (3) & $\pm$0.03 (2) & $\pm$0.10 (1) &   \\
7680 & +0.23 & -0.12 & +0.18 & +0.84 & +0.11 & +0.24 & +0.24 & +0.45 & +0.15 & +0.12 & +0.20 &  7 \\
     & $\pm$0.07 (5) & $\pm$0.04 (4) & $\pm$0.04 (2) & $\pm$0.07 (2) & $\pm$0.10 (1) & $\pm$0.10 (1) & $\pm$0.09 (2) & $\pm$0.10 (1) & $\pm$0.06 (2) & $\pm$0.02 (2) & $\pm$0.04 (2) &   \\
9349 & +0.22 & -0.08 & +0.05 & +0.55 & +0.02 & +0.22 & +0.04 & +0.32 & +0.26 & +0.22 & +0.12 & 13 \\
     & $\pm$0.08 (4) & $\pm$0.09 (4) & $\pm$0.10 (1) & $\pm$0.01 (2) & $\pm$0.01 (2) & $\pm$0.10 (1) & $\pm$0.04 (2) & $\pm$0.10 (1) & $\pm$0.11 (2) & $\pm$0.10 (3) & $\pm$0.10 (1) &   \\
6157 & +0.16 & -0.11 & +0.28 & +0.60 & +0.12 & +0.12 & +0.14 & +0.35 & +0.19 & +0.19 & +0.11 & 12 \\
     & $\pm$0.03 (4) & $\pm$0.04 (4) & $\pm$0.06 (2) & $\pm$0.10 (1) & $\pm$0.00 (2) & $\pm$0.10 (1) & $\pm$0.01 (2) & $\pm$0.10 (1) & $\pm$0.05 (3) & $\pm$0.02 (2) & $\pm$0.10 (1) &   \\
6621 & +0.09 & -0.17 & +0.12 & +0.70 & +0.01 & +0.26 & +0.14 & +0.49 & +0.08 & +0.10 & +0.06 &  9 \\
     & $\pm$0.05 (5) & $\pm$0.06 (4) & $\pm$0.10 (1) & $\pm$0.06 (2) & $\pm$0.10 (1) & $\pm$0.10 (1) & $\pm$0.08 (3) & $\pm$0.10 (1) & $\pm$0.03 (2) & $\pm$0.10 (1) & $\pm$0.06 (2) &   \\
6993 & +0.35 & +0.26 & +0.40 & +0.68 & +0.31 & +0.40 & +0.26 & +0.30 & +0.27 & +0.30 & +0.44 & 15 \\
     & $\pm$0.06 (4) & $\pm$0.04 (3) & $\pm$0.01 (2) & $\pm$0.03 (2) & $\pm$0.05 (3) & $\pm$0.10 (1) & $\pm$0.05 (5) & $\pm$0.07 (2) & $\pm$0.11 (3) & $\pm$0.01 (2) & $\pm$0.03 (2) &   \\
7699 & -0.14 & -0.32 & -0.17 & +0.39 & -0.22 & +0.16 & -0.04 & +0.08 & -0.08 & -0.06 & -0.16 &  9 \\
     & $\pm$0.05 (7) & $\pm$0.07 (3) & $\pm$0.01 (3) & $\pm$0.03 (2) & $\pm$0.08 (2) & $\pm$0.10 (1) & $\pm$0.08 (2) & $\pm$0.06 (2) & $\pm$0.11 (2) & $\pm$0.10 (3) & $\pm$0.00 (2) &   \\
7651 & -0.02 & -0.21 & +0.04 & +0.52 & -0.04 & +0.29 & -0.06 & +0.24 & -0.10 & -0.03 & -0.12 & 10 \\
     & $\pm$0.08 (5) & $\pm$0.04 (5) & $\pm$0.04 (4) & $\pm$0.04 (2) & $\pm$0.06 (3) & $\pm$0.10 (1) & $\pm$0.04 (2) & $\pm$0.11 (2) & $\pm$0.10 (1) & $\pm$0.10 (1) & $\pm$0.06 (2) &   \\
8116 & +0.06 & +0.02 & +0.11 & +0.48 & +0.02 & +0.27 & +0.01 & +0.39 & +0.02 & +0.06 & +0.08 & 11 \\
     & $\pm$0.01 (5) & $\pm$0.05 (4) & $\pm$0.11 (5) & $\pm$0.03 (2) & $\pm$0.07 (3) & $\pm$0.10 (1) & $\pm$0.08 (2) & $\pm$0.10 (1) & $\pm$0.05 (3) & $\pm$0.04 (2) & $\pm$0.06 (2) &   \\
5657 & +0.08 & -0.03 & +0.00 & +0.54 & +0.04 & +0.25 & +0.02 & +0.22 & +0.04 & +0.00 & +0.18 & 14 \\
     & $\pm$0.02 (8) & $\pm$0.07 (5) & $\pm$0.01 (2) & $\pm$0.01 (2) & $\pm$0.01 (2) & $\pm$0.10 (1) & $\pm$0.09 (4) & $\pm$0.01 (2) & $\pm$0.09 (2) & $\pm$0.10 (1) & $\pm$0.01 (2) &   \\
5726 & -0.12 & -0.30 & -0.07 & +0.46 & -0.14 & +0.09 & -0.02 & +0.24 & -0.05 & -0.09 & -0.10 & 12 \\
     & $\pm$0.03 (4) & $\pm$0.06 (5) & $\pm$0.06 (2) & $\pm$0.01 (2) & $\pm$0.10 (1) & $\pm$0.04 (2) & $\pm$0.10 (2) & $\pm$0.10 (1) & $\pm$0.09 (3) & $\pm$0.07 (4) & $\pm$0.10 (1) &   \\
8684 & +0.26 & -0.07 & +0.30 & +0.67 & +0.26 & +0.30 & +0.23 & +0.41 & +0.27 & +0.39 & +0.21 &  8 \\
     & $\pm$0.07 (5) & $\pm$0.10 (5) & $\pm$0.09 (3) & $\pm$0.04 (2) & $\pm$0.02 (3) & $\pm$0.10 (1) & $\pm$0.06 (4) & $\pm$0.10 (1) & $\pm$0.10 (1) & $\pm$0.09 (3) & $\pm$0.05 (3) &   \\
7537 & +0.18 & -0.13 & +0.11 & +0.60 & +0.11 & +0.09 & +0.18 & +0.27 & +0.16 & +0.24 & +0.19 & 10 \\
     & $\pm$0.07 (4) & $\pm$0.02 (5) & $\pm$0.10 (3) & $\pm$0.04 (2) & $\pm$0.06 (3) & $\pm$0.10 (1) & $\pm$0.06 (3) & $\pm$0.04 (2) & $\pm$0.01 (2) & $\pm$0.10 (4) & $\pm$0.10 (1) &   \\
7363 & +0.24 & -0.06 & +0.19 & +0.78 & +0.19 & +0.31 & +0.20 & +0.41 & +0.18 & +0.27 & +0.20 & 11 \\
     & $\pm$0.04 (5) & $\pm$0.03 (5) & $\pm$0.07 (4) & $\pm$0.16 (2) & $\pm$0.08 (3) & $\pm$0.10 (1) & $\pm$0.09 (4) & $\pm$0.05 (2) & $\pm$0.01 (2) & $\pm$0.06 (3) & $\pm$0.02 (2) &   \\
8323 & -0.15 & -0.18 & -0.12 & +0.50 & -0.11 & +0.21 & -0.06 & +0.20 & +0.02 & +0.04 & -0.01 & 11 \\
     & $\pm$0.04 (4) & $\pm$0.03 (3) & $\pm$0.10 (5) & $\pm$0.05 (2) & $\pm$0.01 (2) & $\pm$0.10 (1) & $\pm$0.12 (2) & $\pm$0.06 (2) & $\pm$0.11 (2) & $\pm$0.00 (2) & $\pm$0.10 (1) &   \\
\hline\hline
\end{tabular}

Notes: \\
$^*$~Solar references are from \citet{magg_22}.\\
$^{**}$~The quoted errors are those of the mean, {\it i.e.} the 1$\sigma$ dispersion divided by the square root of the number of used lines (in brackett) to derive that specific abundance.
\end{table*}

Chemical abundances have been obtained via spectral synthesis techniques of suitable atomic lines and molecular features, free from significant blending with other species and/or contamination by telluric absorption and without strong wings. Indeed, although blending from other species is taken into account when using spectral synthesis and telluric correction has been applied, non-perfect modeling and/or correction may leave residuals, hence spectral regions of strong blending and/or contamination have been discarded for chemical analysis. At the R=8,000 resolution of X-shooter,  a few, reasonably clean atomic lines of NaI, MgI, AlI, SiI, KI, CaI, TiI, VI and FeI have been identified  in the H and K bands and they have been used to measure the corresponding element abundances. OH molecular lines and CO bandheads in the H-band have been used to derive O and C abundances, respectively.
From  $^{13}$CO bandheads in the K-band \citep[see e.g.][]{fanelli21} we could also estimate the $^{12}$C/$^{13}$C isotopic ratio. 

Suitable grids of synthetic spectra with stellar parameters as those of the observed stars, with metallicities ranging from [Fe/H]=-1.0~dex to [Fe/H]=+0.5 dex with a step of 0.1 dex, with  some enhancement of [N/Fe] and corresponding depletion of [C/Fe] for a proper computation of the molecular equilibria, and with solar-scaled [X/Fe] values for the other elements, have been computed. We used the radiative transfer code TURBOSPECTRUM \citep{alvarez_98,plez_12}, along with MARCS models atmospheres (\citealt{gustafsson_08}), atomic data from the VALD3 compilation (\citealt{Ryabchikova_15}) and the most recent molecular data from the website of B. Plez\footnote{\url{https://www.lupm.in2p3.fr/users/plez/}}. 

In order to match the observed line profile broadening, the synthetic spectra were convolved with a Gaussian function at the  R$\approx$8,000 X-shooter resolution. This instrumental broadening dominates any other intrinsic broadening, such as macroturbulence and rotation.
For an optimal pixel-to-pixel comparison between the observed and the synthetic spectra, the latter were also resampled to match the pixel size (0.6 $\AA$) of the observed spectra. 
The observed spectra were normalized by using the synthetic spectra as a reference to locally ({\it i.e.} around each line of interest) place the continuum, with an overall uncertainty of $\le$2\%.
Best-fit solutions for the chemical abundances of the various species have been obtained by minimizing the scatter between observed and synthetic spectra around each line of interest and using as a figure of merit both line depth/equivalent width measurements and overall spectral synthesis.
Random errors in the inferred chemical abundances are mostly due to the uncertainties in the placement of the continuum and to the photon noise. 
The values quoted in Table~\ref{tab2} have been thus estimated as the dispersion around the mean abundance divided by the squared root of the number of lines used (normally a few, see Table~\ref{tab2}) to measure each chemical element or an error of 0.1 dex is assumed if only one line is available.

\begin{figure*}
    \centering
    \includegraphics[width=\textwidth]{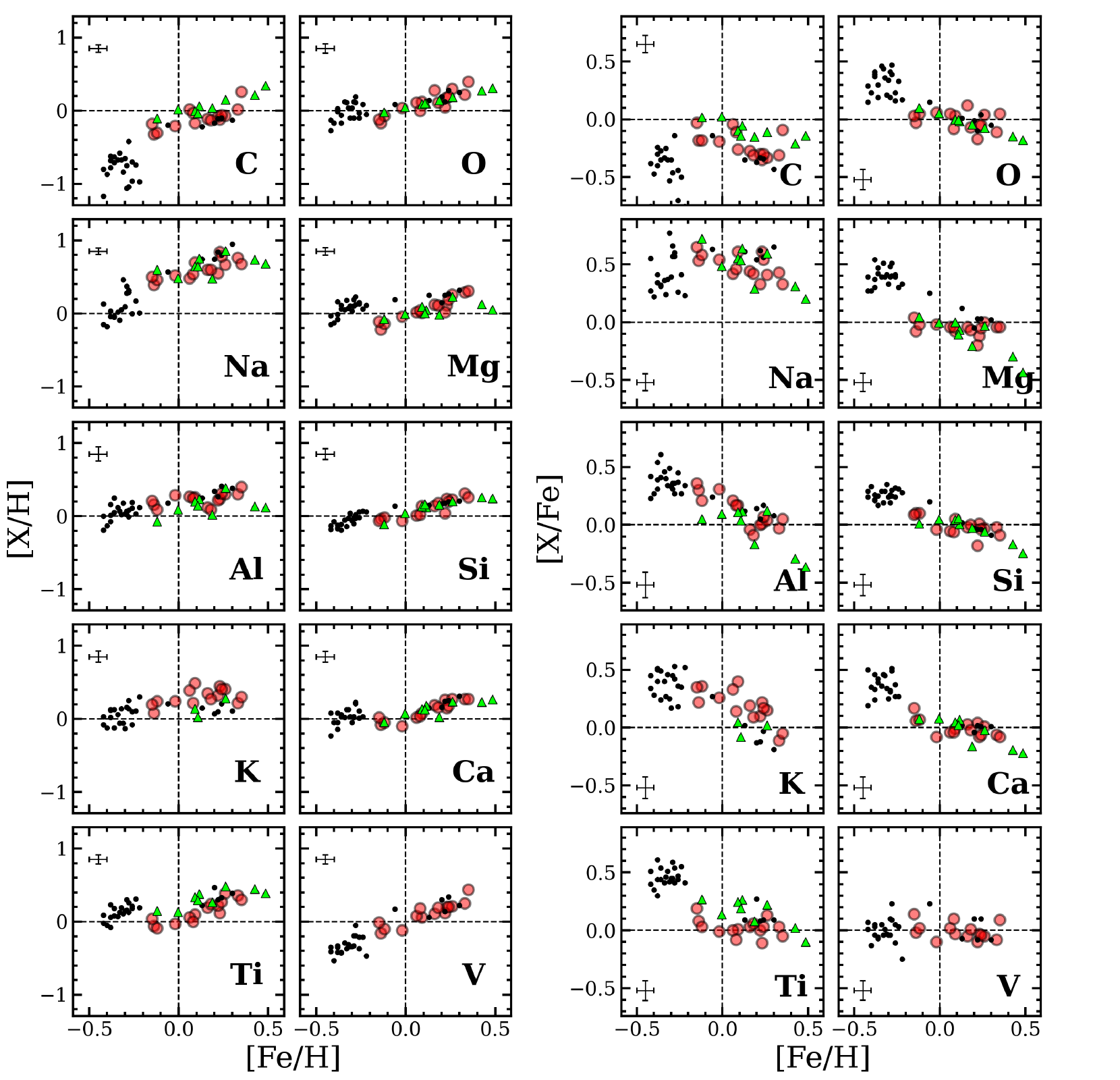}
    \caption{[X/H] abundances (left panels) and [X/Fe] abundance ratios (right panels) of C, O, Na, Mg, Al, Si, K, Ca, Ti, V {\it vs} [Fe/H] for the observed NSC stars (red big dots from this work, and green triangles from \citet{nandakumar25}) and for Liller 1 stars (black dots) from \citet{deimer24}, for comparison. Errorbars are also marked in the left corners of each plot.}
    \label{ratio}
\end{figure*}

\begin{figure}[!h]
    \centering
    \includegraphics[width=\columnwidth]{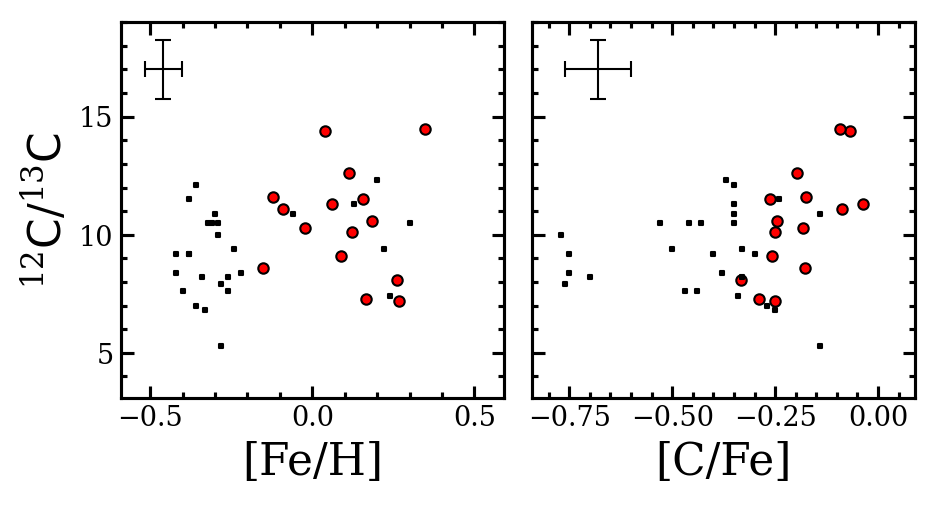}
    \caption{$^{12}$C/$^{13}$C isotopic ratio as a function of [Fe/H] (left panel) and [C/Fe] (right panel) for the observed NSC stars (red big dots) and for Liller 1 stars (black dots) from \citet{deimer24}, for comparison. Errorbars are also marked in the top-left corner of each plot.}
    \label{c12c13}
\end{figure}

\section{Kinematics and orbits}
\label{kin}
For the observed 15 stars heliocentric RVs in the -124 $<$ v$_{\rm hel}$ $<$ +127 km~s$^{-1}$ range, with an average value of -4.6$\pm$16.1 km~s$^{-1}$ and a dispersion of 62.5$\pm$11.4 km~s$^{-1}$ have been inferred. 
Although our estimate of such a velocity dispersion is somewhat affected by low number statistics, it is fully consistent with the velocity dispersion profile obtained by \citet{feldmeier14}, suggests that our sample of candidate NSC stars belongs to a system that should be pressure-supported.
By using these line-of-sight RVs and the corresponding tangential velocities from proper motions by \citet{fritz16}, we also computed the orbital parameters of our stars,  with the main purpose of constraining their confinement to the NSC.
Orbits were computed with \textsc{AGAMA} \citep{AGAMA} in the same non-axisymmetric potential adopted by \citet[][and references therein]{Nieu24}, which includes a triaxial bar/bulge, the NSD, and the NSC. The bar has a clockwise pattern speed $\Omega_p = 40~\mathrm{km\,s^{-1}\,kpc^{-1}}$ and an initial angle of $25^\circ$ with respect
to the Sun--GC line \citep{portail17a,sormani22}. 
We fixed $R_0=8.2$ kpc and performed two different integrations in the bar rotating frame:
{\it i}) an integration over 12 Gyr with a 1 Myr timestep, aimed at characterizing the overall orbital confinement and long-term stability.
{\it ii}) a high-resolution integration over 500 Myr with a 4000 yr timestep, to constrain the orbital morphology.
Most orbits display regular, rosette-like patterns in the x–y plane, while a few show mildly boxy or elongated morphologies depending on their initial phase-space coordinates. The resulting best-fit trajectories of both sets of simulations are confined within $\lesssim 12$ pc from the GC, with $r_{\rm apo}\le 11.6$ pc and $z_{\rm max}\le 7.9$ pc (see Table~\ref{tab1}).

\section{Chemical abundances and abundance ratios} 
\label{chem}

For the observed 15 stars of the NSC, Table~\ref{tab2} lists the inferred chemical abundances and corresponding random errors of Fe, C, O, Na, Mg, Al, Si, K, Ca, Ti, and V  and the $^{12}$C/$^{13}$C isotopic ratio,  by using as solar references those from \citet{magg_22}. 
Fig.~\ref{ratio} shows the computed [X/H] abundances (left panels) and [X/Fe] abundance ratios (right panels) as a function of [Fe/H], with typical errors on individual element abundances of $\le$0.1~dex (see Table~\ref{tab2}). 
The inferred distribution of [Fe/H] values for the observed 15 stars is quite broad, spanning about 0.5 dex, from -0.15 to +0.35 dex, and with an average value of +0.12$\pm$0.04 dex and a 1$\sigma $ dispersion of 0.17$\pm$0.03 dex, significantly larger than the error of the mean and also somewhat larger than the typical $<$0.1 dex measurement error.
Such a metallicity spread suggests that the NSC is a stellar system more complex than a genuine globular cluster. 
The  [X/Fe] abundance ratios of O, Mg, Si, Ca, Ti and V are about solar-scaled (within $\pm$0.1 dex), while [Na/Fe] is enhanced with respect to the solar-scaled value. [Al/Fe] and [K/Fe] distributions show some trend with metallicity, with more metal-poor stars being  somewhat enhanced, while the more metal-rich ones being about solar-scaled.
We do not find any significant dispersion of Na and/or O and consequently no evidence of a Na-O anti-correlation, typically observed in genuine globulars \citep[e.g.][]{caretta_09}.

These [X/Fe] {\it vs} [Fe/H] distributions have been compared to the ones of Liller~1, a complex stellar system of the bulge hosting multi-age and multi-metallicity stellar populations \citep{ferraro_21}, for which a similar chemical analysis from  X-shooter spectroscopy \citep{deimer24,fanelli24} has been performed. 
We find  a nice agreement between the abundance patterns of the NSC stars and the metal-rich population of Liller~1, and more generally with the metal-rich stellar populations of the bulge and other complex stellar systems like Terzan~5 \citep[see e.g.,][and references therein]{rich12,Origlia_11}.
In particular, the [$\alpha$/Fe] abundance ratios consistent with the solar-scaled values or at most with a mild enhancement, indicates that these metal-rich stars formed from a gas enriched by both type II and Type~I SNe, at variance with the 
metal-poor components of the bulge stellar populations that show a significant [$\alpha$/Fe] enhancement and likely formed more quickly from a gas mostly enriched by type II SNe.

Our chemical analysis of the NSC stars is also in nice agreement with the study of \citet{nandakumar25} for the chemical elements in common. 

Some [C/Fe] depletion (see Fig.~\ref{ratio}) with respect to the solar-scaled value, low $^{12}$C/$^{13}$C isotopic ratios between 7 and 15 (see Table~\ref{tab2} and Fig.~\ref{c12c13}) with a typical uncertainty of $\pm$1 and some correlation between $^{12}$C/$^{13}$C and [C/Fe] for all the observed stars have been also inferred from our study. Such a carbon depletion is typically measured in luminous giants, regardless of their metallicity, indicating that mixing and extra-mixing processes should be at work during the evolution along the RGB.

\section{Discussion and conclusions}
\label{conc}
Our kinematic and chemical study of an homogeneous sample of 15 giant stars, all located within 0.5 and 1.6 pc from the GC, {\it i.e.} well within the half-light radius of the NSC, with line-of-sight RVs and proper motions also consistent with a NSC membership, and color-magnitude diagram consistent with being very reddened giants, has some statistical significance to attempt a first characterization of the old stellar content of the NSC in terms of population properties. 

With the purpose of evaluating whether the observed distribution of $[\alpha/$Fe] (where $\alpha$ is the average of the O, Mg, Si, Ca abundances) {\it vs} [Fe/H] shows some statistical evidence of multiple sub-components, we performed a Gaussian Mixture Model (GMM) analysis, as done  by \citealt{fanelli24} on Liller~1. Both the Bayesian and Akaike Information criteria favour a single population model. We caution, however, that with only 15 stars this analysis has very limited statistical power. Hence, we also implemented a 2D ([Fe/H], [$\alpha$/Fe]) hierarchical Bayesian model, using the {\tt dynesty} dynamic nested sampling package and accounting for the full covariance matrix in the error budget of each star. According to this test, the single-population model is moderately favoured on the Jeffreys scale, although the existence of a more complex structuring, undetectable with the current sample size, cannot be ruled out.

\begin{figure}
    \centering
    \includegraphics[width=\columnwidth]{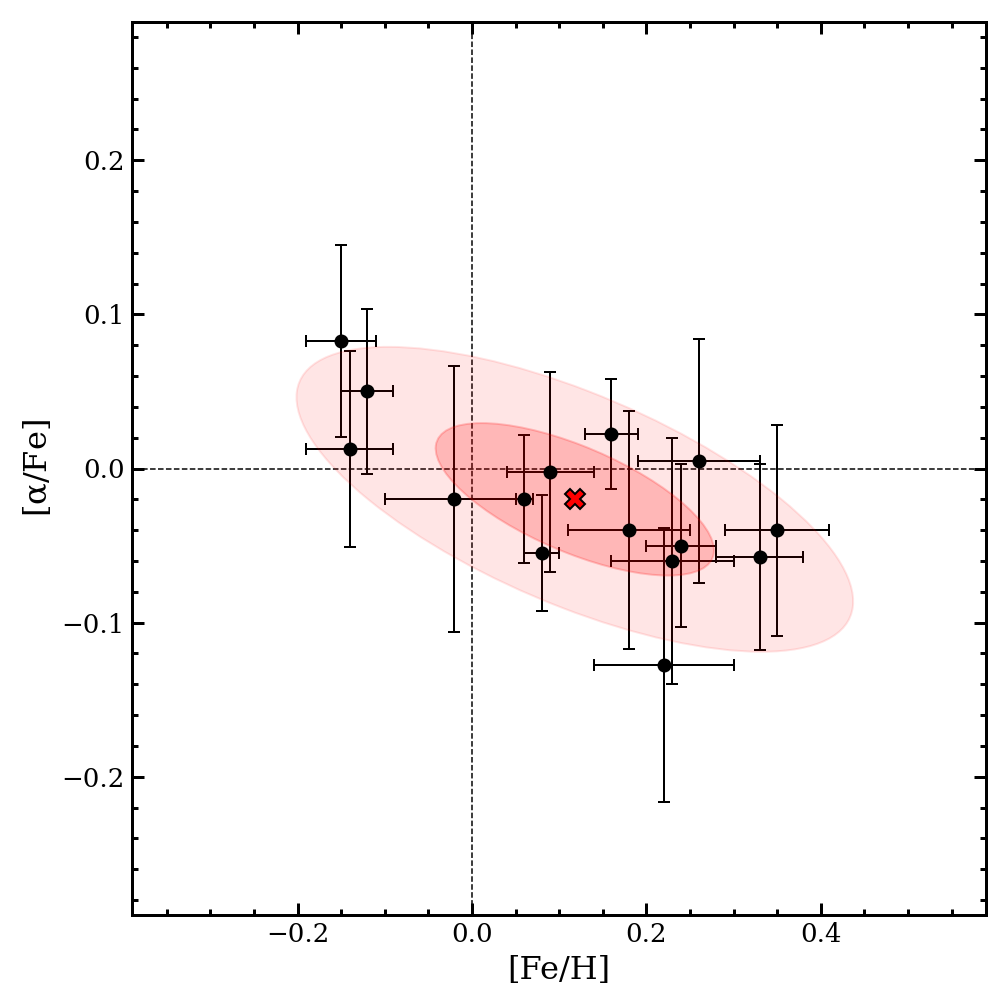}
    \caption{Average $[\alpha/$Fe] {\it vs} [Fe/H] and associated errorbars for the observed NSC stars (black dots), where $\alpha$ is the average of the O, Mg, Si and Ca abundances. The Gaussian Mixture Model (GMM) best-fit is superimposed, the red cross indicating the centroid (mean) of the identified component, while the concentric shaded regions being the $1\sigma$ and $2\sigma$ confidence ellipses.}
    \label{alpha}
\end{figure}

The overall high metallicity slightly above solar, on average, and with significant spread, together with the rather low (if any) level of [$\alpha$/Fe] enhancement (see Fig.~\ref{alpha}) of such a population  suggests that the gas from which it formed may have [self-]enriched with both type II and type I SNe ejecta on a somewhat prolonged timescale.
Moreover, the kinematic confinement in the central parsecs and the striking chemical similarity with the metal-rich stellar populations of the Galactic bulge, as well as  of other complex stellar systems there located, like Terzan~5 and Liller~1, strongly favor an in-situ formation and evolution of the NSC.

In the coming years we have the perspective of making a true quantum step in characterizing the stellar content of the nuclear region of the Milky Way, thanks to the Galactic survey in {\it guarantee time} with the new multi-object spectrograph MOONS \citep{moons20} at the VLT.
More specifically, as a part of that survey, there will be an optimized sampling of the stellar populations in the central hundred parsecs, by acquiring medium-high resolution NIR spectra of a dozen thousand stars distributed in the nuclear disk/bar/bulge and of a few hundreds stars in the NSC. 
Such a high quality, homogeneous dataset will provide the necessary statistical ground to perform robust chemo-dynamical population studies, in order to properly identify and characterize possible sub-structures and their mutual interactions, as well as their link with the inner Galaxy.

\begin{acknowledgements}
LO and CF acknowledge the financial support by INAF within the ELT-ANDES project. CF acknowledges support by the PRIN INAF 2023 grant ObFu N2RED (PI: C. Fanelli). NR acknowledges support from the Swedish Research Council (grant No. 2023-04744).
\end{acknowledgements}

\bibliographystyle{aa} 
\bibliography{mybib} 
\end{document}